# Risks and Benefits of Using a Commercially Available Ventricular Assist Device for Failing Fontan Cavopulmonary Support: A Modeling Investigation

Masoud Farahmand, Minoo N. Kavarana, Ethan O. Kung*

*Abstract*— Fontan patients often develop circulatory failure and are in desperate need of a therapeutic solution. A blood pump surgically placed in the cavopulmonary pathway can substitute the function of the absent sub-pulmonary ventricle by generating a mild pressure boost. However, there is currently no commercially available device designed for the cavopulmonary application; and the risks and benefits of implanting a ventricular assist device (VAD) originally designed for the left ventricular application on the right circulation of failing Fontan patients is not yet clear. Moreover, further research is needed to compare the hemodynamics between the two clinically-considered surgical configurations for cavopulmonary assist, with Full and IVC Support corresponding to the entire venous return or only the inferior venous return, respectively, being routed through the VAD. In this study, we used a numerical model of the failing Fontan physiology to evaluate the Fontan hemodynamic response to a left VAD during the IVC and Full support scenarios. We observed that during Full support the VAD improved the cardiac output while maintaining blood pressures within safe ranges, and lowered the IVC pressure to <15mmHg; however, we found a potential risk of lung damage at higher pump speeds due to the excessive pulmonary pressure elevation. IVC support, on the other hand, did not benefit the hemodynamics in the patient cases simulated, resulting in the superior vena cava pressure increasing to an unsafe level of >20 mmHg. The findings in this study may be helpful to surgeons for recognizing the risks of a cavopulmonary VAD and developing coherent clinical strategies for the implementation of cavopulmonary support.

## I. INTRODUCTION

Over the past 50 years, Fontan operation has been the most common palliative care for patients with single ventricle defect. Currently 50,000-70,000 patients exist worldwide that have been able to survive into adulthood because of the Fontan operation [1]. Ohuchi et al. [2] studied the hemodynamics in nearly 500 Fontan patients, separating them into early (0.5-5 years postoperative) and late (≥15 years postoperative) groups; the age range for the early and late groups were 9±6 years and 23±7 years, respectively; they showed that the long-term outcomes of the Fontan operation is associated with high mortality. A significant number of these patients will eventually develop circulatory failure as a result of the poor hemodynamics following the Fontan procedure [3], [4]. Based on a 10 year follow up study, 1 out of 4 children after the Fontan operation are not expected to survive [4]. Previous clinical measurements [4]–[6] have recorded pulmonary arterial hypotension and high caval pressure (also known as the "Fontan paradox") in single ventricle patients after the Fontan procedure. Chronic high caval pressure is found to be the main determinant of many Fontan-related diseases and circulatory failure [7]. Failing stage in a Fontan circulation manifests itself in a cascade of the pathological consequences such as protein-losing enteropathy, hepatic failure, and gut dysfunction [8].

Most failing Fontan patients are unable to receive heart transplantation either because they are poor candidates as a result of the several previous surgeries and multiple organ dysfunctions, or because of the unavailability of matching organ donors; these patients are in need of an immediate therapeutic solution. The use of ventricular assist device (VAD) for systemic circulatory support (left support) [9], [10] or cavopulmonary support (right support) [11]–[13] has been proposed for treatment of Fontan failure. This study focuses on the cavopulmonary support scenario where a mechanical assist device propels the venous return forward into the lungs as an attempt to reverse the Fontan paradox. Conceptually, a pump properly implemented in the cavopulmonary pathway could generate a slight pressure boost that benefits the hemodynamics by lowering the caval pressure and augmenting the cardiac output. Earlier studies either have suggested implementing the cavopulmonary device to directly pump the inferior vena cava (IVC) flow to the pulmonary artery ("IVC assisted configuration") [13]–[16] or proposed the "Fully assisted configuration" to pump the entire venous return to the pulmonary artery [11], [17] (Fig. 1). Previously, Molfetta et al. [16] showed that an ideal cavopulmonary pump could promote the cardiac output by 34% during IVC support. To our knowledge, there has been no study that directly assesses how the hemodynamics compare between these two surgical

Manuscript received December 05, 2018, revised March 11, 2019, and accepted April 10, 2019. This work was supported by Clemson University, an award from the American Heart association and The Children's Heart Foundation (16SDG29850012) and an award from the National Science Foundation (1749017).

M. Farahmand is with Clemson University, Clemson, SC, USA. M. N. Kavarana is with Medical University of South Carolina, Charleston, SC, USA. *E. O. Kung is with Clemson University, Clemson, SC, USA. (correspondence email: ekung@clemson.edu)

configurations for the installation of the cavopulmonary assist device.

A concerted work is in progress to develop a cavopulmonary assist device as a bridge to cardiac transplantation or for long-term support. Researchers have suggested several prototypes for powering the Fontan circulation [14], [18]–[21]. While these prototypes are designed to be compatible with a low-pressure operating environment, they are not commercially available and thus not employed in the clinical management of single ventricle patients. As such, a relevant clinical question is whether an existing VAD originally designed for the left side application can be used as right-side support in a failing Fontan patient.

Therefore, the objectives of this study are 1) to identify the possible risks and benefits of the off-label use of the Jarvik 2000 (or another left VAD with similar mechanical performance) for IVC and Full support in adult failing Fontan patients, and 2) to assess how the hemodynamics compare between the IVC and Fully assisted surgical configurations for installation of the cavopulmonary assist device. We hypothesized that it is possible to use a left VAD for the cavopulmonary application and to palliate the failing Fontan condition albeit with potential risks. To test this hypothesis, we first performed a sensitivity analysis on a lumped parameter network (LPN) model of the healthy adult Fontan physiology to attain a quantitative understanding how different factors affect the Fontan circulation; next we constructed a failing Fontan circulation model and simulated the IVC and Fully assisted surgical configurations to determine hemodynamic aberrations of each configuration at different pump speeds.

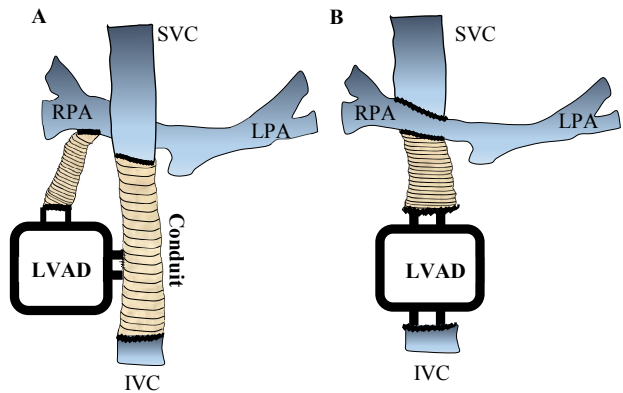

Fig. 1. Schematic representation of the surgical configurations for the cavopulmonary assist installation. **(A)** Fully assisted configuration (Full support) and **(B)** IVC assisted configuration (IVC support). LPA: Left pulmonary artery, RPA: Right pulmonary artery SVC: Superior vena cava, IVC: Inferior vena cava

## II. MATERIAL AND METHODS

### A. Physiology model of the healthy Fontan circulation (baseline)

A closed-loop LPN was employed to model the Fontan physiology (Fig. 2). A full description of the modelling protocol has been described in our previous work [22]. Briefly, the LPN can adjust model parameters such as cardiac function, vascular resistances, vascular compliances, and respiratory parameters to simulate the hemodynamics of a Fontan patient. The transmyocardial pressure ($P_{sv}$) generated by the single ventricle drives the flow in the circuit and is defined to be correlated with the ventricular elastance and volume. The ventricular time-dependent elastance "$E(t)$" is defined as a function representing ventricular filling and contraction.

$$P_{sv}(V_{sv}, t) = E(t)(V_{sv} - V_{sv0}) \quad (1)$$
$$E(t) = E_{max} En\left(\frac{0.3}{t_{svs}} t\right) + E_{offset} \quad (2)$$

$t$ is the time point in the cardiac cycle. The systolic period is given by $t_{svs}$, where $V_{sv0}$ and $V_{sv}$ are the ventricular reference volume and ventricular volume, respectively. $En$ is the normalized elastance function. $E_{max}$ is related to the contractility and ventricular ejection fraction. $E_{offset}$ reflects ventricular filling. The ventricular compliance is correlated with ventricular filling and thus $E_{offset}$ is also considered as a measure of the ventricular compliance.

In this study, we tuned the LPN models according to [22] to represent two healthy adult example Fontan patients with a small and large body surface area (BSA) under resting condition. The small patient has weight=50kg and height=150cm and the large patient has weight=75kg and height=180cm.

### B. Sensitivity analysis

Most of the single ventricle patients following the Fontan operation display signs of ventricular diastolic dysfunction and elevated pulmonary vascular resistance [23]–[25]. To determine how ventricular diastolic dysfunction and pulmonary vascular resistance affect hemodynamic parameters such as cardiac output, atrial pressure, and IVC pressure in a failing Fontan circulation, we performed a sensitivity analysis using LPN simulations where in each example patient we decreased the ventricular compliance by an amount from 0 to 75% and increased the pulmonary vascular resistance by an amount from 0 to 100% of their baseline values with a step change of 25%. We maintained the ventricular contractility at its baseline value because previous studies have shown that ejection fraction is mostly preserved in the majority of failing and healthy Fontan patients [24], [26], [27].

### C. Physiology model of the failing Fontan circulation

A failing circulation is characterized by high pulmonary vascular resistance and low cardiac output [7]. Increased pulmonary vascular resistance is partly attributable to low pulmonary pressure and attenuated pulmonary vascular pulsatility in Fontan patients [25]. Clinical reports show that on average in a failing Fontan circulation, pulmonary vascular resistance is elevated by 68% [7] and cardiac output is reduced by 40% [7], [28] compared to the healthy Fontan circulation.

Moreover, Hebson et al. [5] reported that adult failing Fontan patients had lower systemic vascular resistance in comparison with asymptotic Fontan patients, possibly due to the autonomic control maintaining the total vascular resistance. A previous report showed that 73% of the Fontan patients also suffer from a decrease in ventricular compliance (resulting in ventricular diastolic dysfunction) that initiates early after the Fontan procedure possibly as a result of preload depletion [24].

To simulate the failing Fontan circulation based on these clinical findings, we first increased the pulmonary vascular resistance by 68% from the baseline. Then, we reduced the systemic vascular resistance to compensate for the elevated pulmonary vascular resistance such that the total vascular resistance is maintained. Finally, we decreased the ventricular compliance from the baseline to obtain a 40% reduction in the cardiac output emulating ventricular diastolic dysfunction.

### D. Model of cavopulmonary support

Using the failing Fontan LPN, we modelled the IVC and Fully assisted cavopulmonary support configurations as shown in Fig. 2. Two thirds of the left VADs implanted in patients in the USA are axial-flow VADs [29], many with similar hydraulic characteristics curves as the Jarvik 2000 (Jarvik Heart, Inc., New York, NY, USA). The performance range of the Jarvik 2000 can encompass the characteristic curves of several commonly used VADs operating at their lowest rotor speeds (Fig. 3A) [30], [31]. Therefore, we modelled the Jarvik 2000 as a representative VAD in this study.

The off-the-shelf operating range of the Jarvik pump is 8000-12000rpm. We modified the off-the-shelf controller to achieve lower pump speeds down to 4400rpm, then performed a series of physical flow experiments to characterize the blood pump (Fig. 3B). The details of the flow experiments are included in the supplementary materials. The pump model is presented in the form of a quadratic equation [32], [33].

$$\Delta P = A\,Q^2 + BQ + C \quad (3)$$

Where $\Delta P$ is the pressure head across the pump and $Q$ is the flowrate through the pump ranging from 0 to 6 L.min$^{-1}$. $A$, $B$, and $C$ are constants dependent on the pump speed (Table. 1).

### III. RESULTS

#### A. Sensitivity analysis

We quantified the sensitivity of the Fontan hemodynamics to variations in the pulmonary vascular resistance and ventricular compliance (Fig. 4). Both example patients we simulated had a similar response to the sensitivity analysis. While the cardiac output and atrial pressure were affected by both pulmonary vascular resistance and ventricular compliance, they were more sensitive to the ventricular compliance. For every 25% decrement in the compliance, the cardiac output decreased by ~11% and atrial pressure increased by ~24%. For every 25% increment in the pulmonary vascular resistance, the cardiac output decreased by ~3%, and atrial pressure decreased by <2.5%. Moreover, the slope of the lines at different levels of ventricular compliance indicated that the atrial pressure became slightly more sensitive to the pulmonary vascular resistance at lower ventricular compliances. The IVC pressure is sensitive to changes in both the pulmonary vascular resistance and the ventricular compliance.

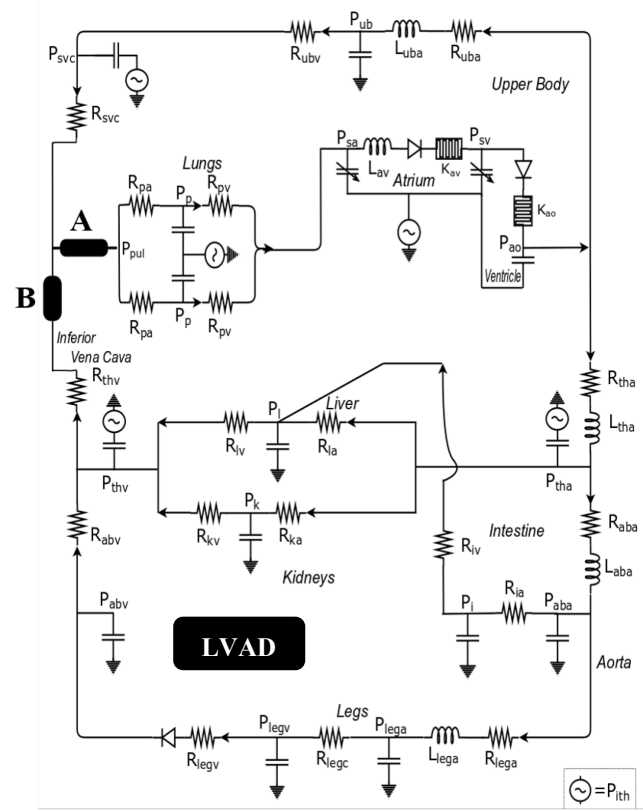

Fig. 2. LPN model of the Fontan circulation coupled with the VAD describing either the (**A**) Full Support or (**B**) IVC Support scenario. Pressure points ($P_{sub}$) are labelled on the diagram. $L_{sub}$ and $R_{sub}$ are inductor and resistor components, respectively. [22]

TABLE 1
Coefficients for the Jarvik 2000 blood pump performance characteristic model at different rotor speeds.

| Rotor speed (rpm) | Coefficients | | |
|---|---|---|---|
| | A | B | C |
| 4400 | -0.002356 | -0.1267 | 20.55 |
| 5500 | -0.003486 | -0.0711 | 30.83 |
| 7000 | 0.01323 | -0.8355 | 56.91 |
| 8500 | 0.0106 | -0.854 | 79.86 |
| 9500 | 0.02024 | -1.511 | 107.2 |

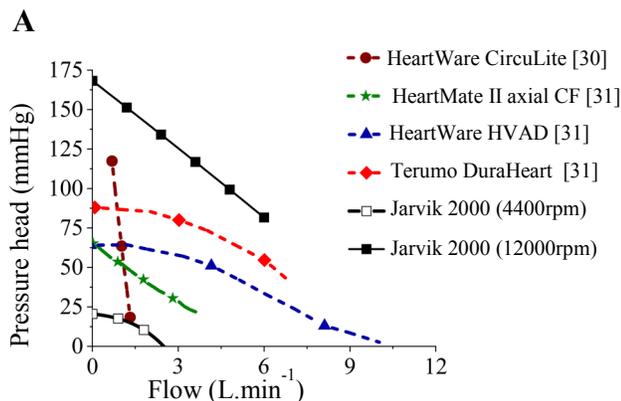
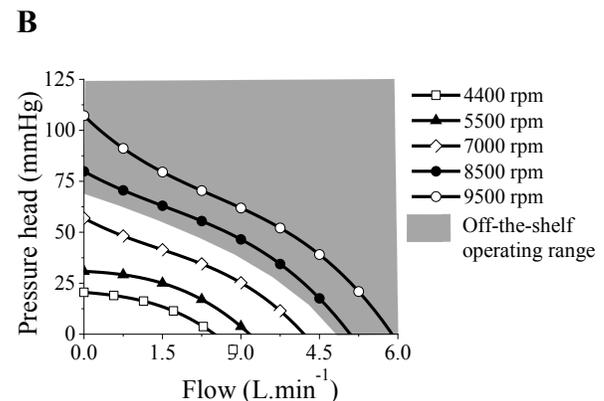

Fig. 3. Blood pump characteristic curves. (**A**) Comparison of head curves of several commonly used left VADs operating at their lowest rotor speeds vs the operating range of the Jarvik 2000, and (**B**) Performance curves of the Jarvik 2000 blood pump at different rotor speeds.

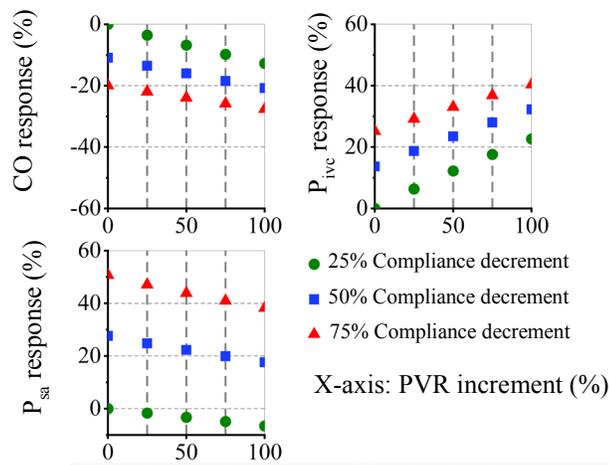

Fig. 4. Hemodynamic response of the large simulated patient to the variations in the pulmonary vascular resistance (PVR) and ventricular compliance. CO: Cardiac output, $P_{ivc}$: IVC pressure and $P_{sa}$: Atrial pressure.

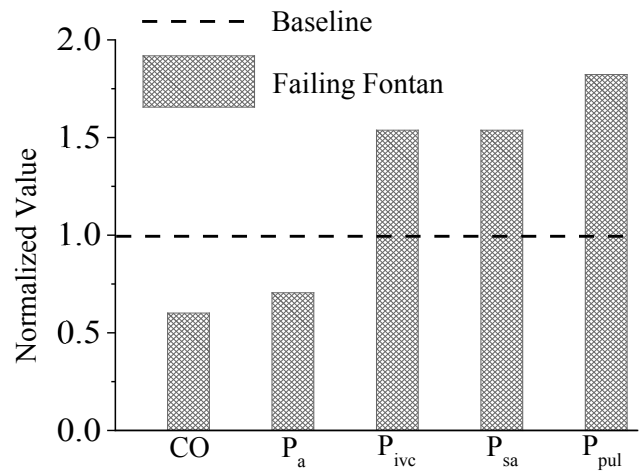

Fig. 5. Failing Fontan hemodynamics in comparison with the baseline. The averaged hemodynamic data of two example failing Fontan patients were normalized with respect to the baseline. Cardiac output (CO), arterial pressure ($P_a$), IVC pressure ($P_{ivc}$), pulmonary pressure ($P_{pul}$) and atrial pressure ($P_{sa}$) are presented.

### B. Failing Fontan physiology

In our simulations, the hemodynamic parameters of a failing circulation were significantly different from those of the healthy Fontan patients (Table. 2 and Fig. 5). Compared to the baseline, the arterial pressure in the failing circulation was significantly lower (~30%); the clinical data reported by Cavalcanti et al. [28] and Egbe et al. [6] similarly showed a 23% decrease in arterial pressure. The simulations indicated that the IVC pressure elevated on average by ~50% (6.5mmHg) in the failing Fontan patients comparing to the baseline, also consistent with previous reports [5], [6], [28]. The atrial pressure substantially increased in the failing Fontan models by an average of ~81 %.

TABLE 2
Mean values of the hemodynamic parameters from simulations of the two example Fontan patients at baseline and during Fontan failure.

| Parameters | Baseline / Failing | |
|---|---|---|
|  | Small patient | Large patient |
| Height (cm) | 150 | 180 |
| Weight (kg) | 50 | 75 |
| BSA (m$^2$) | 1.44 | 1.94 |
| Cardiac output (L.min$^{-1}$) | 3.56 / 2.14 | 4.71 / 2.86 |
| Arterial pressure (mmHg) | 89.44 / 63.10 | 87.64 / 61.69 |
| Atrial pressure (mmHg) | 8.02 / 14.62 | 7.46 / 13.45 |
| Pulmonary pressure (mmHg) | 12.29 / 18.90 | 11.66 / 17.88 |
| IVC pressure (mmHg) | 12.29 / 18.90 | 11.66 / 17.88 |

IVC: Inferior vena cava

### C. Cavopulmonary support for the failing Fontan circulation (Fig. 6)

#### 1) Fully assisted configuration

The presence of the VAD increased the cardiac output and favourably decreased the IVC pressure across all pump speeds for the small patient but only above 5000rpm for the large patient. For the large patient with rpm <5000, the cardiac output and the IVC pressure deteriorated and changed by ~ -7.5% and ~+5.6%, respectively.

For both patients, the cavopulmonary VAD generally resulted in the reduction of the superior vena cava (SVC) pressure. For every 1000rpm increase in the pump speed, the IVC pressure decreased by ~20% and the cardiac output improved by ~ 10% on average. Higher pump speeds (small patient: >~5500rpm, large patient: >~6800rpm) led to the elevation of the pulmonary pressure level to >25mmHg, which is above safe physiological range [34]. Overall, the rpm values for improving cardiac output while maintaining pressures within safe physiological range was between 5000-6800 for the large patient and 4400 to 5500 rpm for the small patient.

#### 2) IVC assisted configuration

While the IVC support augmented the cardiac output and led to a favourable decrease in the IVC pressure, the SVC pressure increased to above 20mmHg even at the lowest pump speed (4400rpm) for both patients. The IVC pressure was highly sensitive to the pump speed, continually decreasing with increasing pump speed.

### IV. DISCUSSION

The Fontan procedure is currently the standard-of-care final procedure for the palliation of single ventricle defects. However, Fontan patients are predisposed to circulatory failure as a result of the abnormal hemodynamics. This study aims to evaluate the risks and benefits of the off-label use of a left VAD

TBME-01987-2018

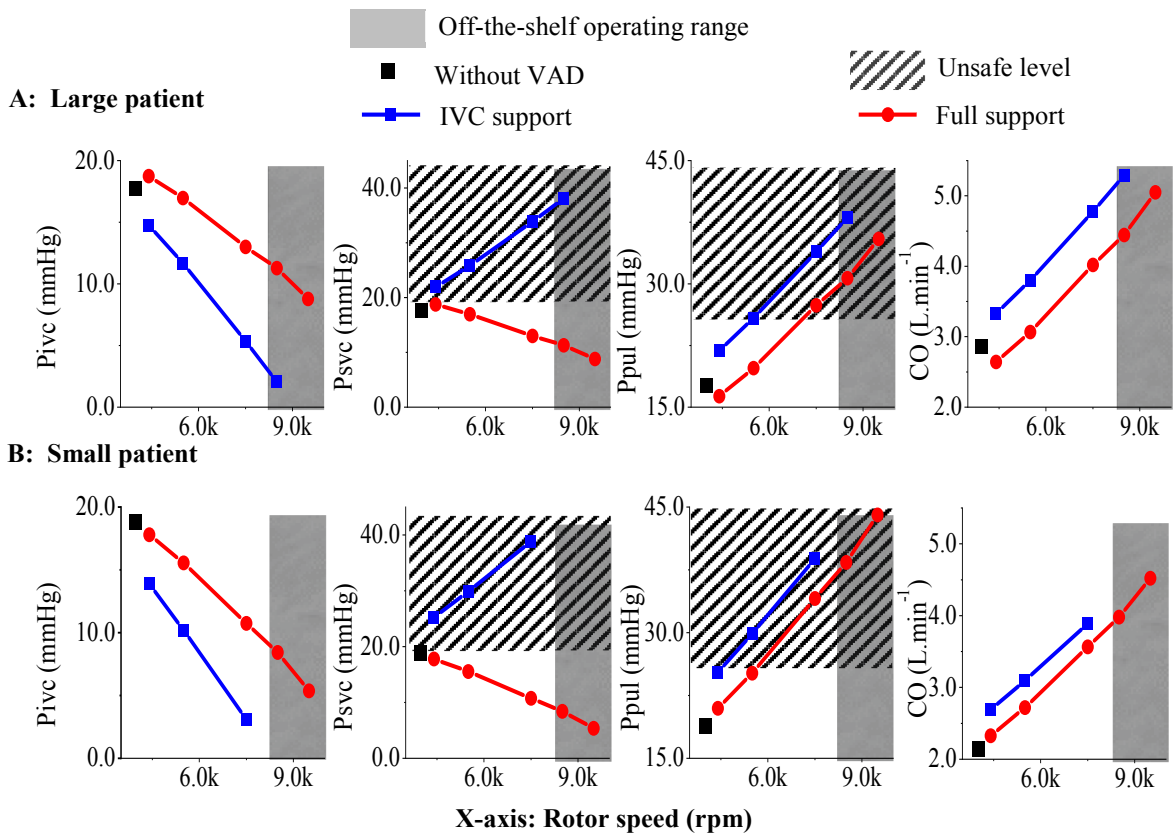

Fig. 6. Comparison of hemodynamic parameters during Full and IVC support at different rotor speeds. Simulation results of the (**A**) large patient and (**B**) small patient. Cardiac output (CO), SVC pressure ($P_{svc}$), IVC pressure ($P_{ivc}$), pulmonary pressure ($P_{pul}$).

during Full and IVC support in failing Fontan patients with diastolic dysfunction. Through simulations, we showed that a proper implementation of an arterial pump in the right circulation could benefit the Fontan hemodynamics. We constructed the failing Fontan model employing clinical data available in the literature and found that the failing circulation model to be in general agreement with previous reports [16], [18], [35]. Our model sensitivity analysis showed the association of the hemodynamic parameters (e.g. cardiac output, IVC pressure, atrial pressure) with pulmonary vascular resistance and ventricular diastolic dysfunction. The results of this analysis suggested that the significant elevation of IVC pressure in a failing circulation reported in previous clinical studies [5], [28] is equally attributable to the high pulmonary vascular resistance as well as elevated downstream pressure related to ventricular compliance reduction. Furthermore, deterioration of ventricular compliance had significant implications on the atrial pressure, cardiac output, and central venous pressure; thus, it is important to include ventricular diastolic dysfunction in failing Fontan circulation models. The sensitivity analysis indicated that changes in the ventricular compliance and pulmonary vascular resistance had considerable and minimal effects on atrial pressure, respectively. This implies that elevated atrial pressure may be an indication of ventricular diastolic dysfunction in failing Fontan patients. The sensitivity analysis also showed that ventricular diastolic dysfunction and elevated pulmonary vascular resistance both decreased the cardiac output. Cardiac output reduction and consequently further preload starvation most likely will result in additional deterioration of the ventricular diastolic function [23], [36] leading to a positive feedback loop and eventually cardiac failure in failing Fontan patients. Therefore, augmentation of the cardiac output is a crucial factor for the survival of these patients.

The utilization of the right-side circulatory device to promote the hemodynamics of a failing Fontan is appealing; however, implanting the cavopulmonary device on the right circulation is surgically complex. Implementation of the pump in a failing Fontan requires additional invasive surgeries for taking down the venae cavae for cannulation. From a surgical point of view, the Full assist configuration is relatively more invasive because of the additional steps necessary for detaching and anastomosing both the SVC and IVC.

However, our simulations showed that IVC support had limitations. While surgical implementation of IVC assist is perhaps less complex, the model revealed that the pressure generated by the pump even at the lowest speed was excessively high (7mmHg at 1.66 L/min of flow) and resulted in elevation of the SVC and pulmonary pressure to above 22mmHg and out of acceptable physiological ranges. Cerebral venous thrombosis, a result of elevated central venous pressure, is frequent after the Fontan operation (occurring in 21% of patients) [37]–[39]. Precipitous rise of the SVC pressure due to an IVC assist pump transmitted to the cerebral venous circulation could further predispose the patient to thrombosis.

Shimazu et al. [15] reported that IVC assist is more suitable for Fontan cavopulmonary support. They modelled a similar size example patient with weight=75 kg and BSA=1.9 m$^2$. The disparity between our conclusions is likely due to the differences in the pump models employed and the cardiac function prescribed in the physiology simulations. The pump model that they employed was more compatible with the low-pressure environment. This pump, although not currently commercially available, was capable of generating smaller pressure boost at low flow rates (resulting in lower SVC pressure). Additionally, their example patient had a normal ventricular diastolic function and consequently higher IVC flow and cardiac output. Nonetheless, their study also confirmed that the elevation of the SVC pressure by the presence of the pump is a major drawback of the IVC assist configuration.

Risk of venous collapse is another important concern with the IVC assist configuration while using a left VAD. Rieme et al. [13] observed the intermittent collapse of the IVC and subsequent decrease in the IVC flow in animals after implementing a HeartMate II in the IVC. Our simulation results also implied the possibility of venous collapse at the inlet of the pump due to the very low IVC pressure at high pump speeds. On the other hand, for the fully assisted configuration the caval pressure remained positive across all pump speeds we examined; this result is consistent with the report by Gandolfo et al. [17] where no venous collapse was observed after implanting the Jarvik 2000 on the right circulation of four sheep using the Full Support surgical configuration.

Although the risk for venous collapse was lower in Full support, the simulations showed that over-filling of the Jarvik pump at low pump speeds was a major shortcoming of this configuration. In the large patient and for pump speeds <5000rpm, the blood pump obstructed the flow and resulted in a reduction of the cardiac output and undesirable increase of the IVC and SVC pressure. However, in the small patient no pump over-filling was observed due to the lower venous return flow rates.

During Full support, we were able to identify a range of the pump speed for each example patient that resulted in improved hemodynamics. However, the optimal range was outside of the off-the-shelf range of the Jarvik 2000 (i.e. 8000-12000rpm). While maintaining all the pressure in the safe physiologic range, the cardiac output increased by up to 30% at 5500rpm and 6800rpm for the small and large patients, respectively. The optimal pump speeds depended on the specific hemodynamics of the failing patient and performance of the pump. Therefore, subject-specific modelling is important for identifying suitable pump speeds while investigating cavopulmonary assist.

*Limitations*

Currently, the cavopulmonary device is not employed in the clinical management of Fontan patients, therefore the results of this study cannot be validated against clinical measurements. Fluctuation in the rotational speed of the pump as a result of pulsatile flow has been observed in our previous experimental study [33]; while the LPN model is not capable of capturing these fluctuations, since the pulsatility in the pulmonary flow is low, these fluctuations should not result in significant error in the simulated hemodynamic outcomes. Another limitation of the current study is that the LPN model is not able to resolve vessel wall shear stress, which is known to impact endothelial function and pulmonary vascular resistance in Fontan patients. Finally, the numerical model is tuned based on clinical data from adult patients [22], therefore the results should not be expanded to pediatric patients.

## V. CONCLUSION

The computational model we presented suggested that a cavopulmonary VAD can benefit failing Fontan patients and that IVC assisted configuration is associated with higher risks compared to Full assist. The proper adjustment of a cavopulmonary device depends on the device characteristics and hemodynamic state of the patient; we therefore recommend subject-specific modelling for identifying the pump speed range that would benefit the patient. Future studies should focus on designing a blood pump that has the performance requirements for cavopulmonary application. We showed that, theoretically, a modified left VAD (Jarvik 2000) implemented in the right circulation can help failing Fontan patients. However, further research is needed to examine the risks associated with the modification of a left VAD to achieve lower pump speeds since such modifications can affect the reliability and stability of the device. Other methods for reducing the pressure generated by the pump such as trimming of the rotor blades [17] can also be further investigated in future studies.

ACKNOWLEDGMENT

We thank J. Teal and J. Triolo at Jarvik Heart Inc. for the equipment loan, Himanshu Deshmukh and Nathan McDowell for manuscript editing assistance. The authors acknowledge the work of members of our laboratory. The funding sponsors of this work have no involvement in the study design, analysis and interpretation of data, in the writing of the manuscript, and in the decision to submit the manuscript for publication.